\documentclass{iopconfser}

\usepackage{amsmath}
\usepackage{amssymb}
\usepackage{graphicx}
\usepackage{enumitem}
\usepackage{lmodern}
\usepackage{multirow}
\usepackage{supertabular}
\usepackage{xspace}

\usepackage{subcaption}

\newcommand{\fbinv} {\mbox{\ensuremath{\,\text{fb}^{-1}}}\xspace}
\newcommand{\GeV}{\ensuremath{\,\mathrm{Ge\hspace{-.08em}V}}\xspace}
\newcommand{\pT}{\ensuremath{p_{\mathrm{T}}}\xspace}

\begin{document}

\title{Line Segment Tracking: Improving the Phase 2 CMS High Level Trigger Tracking with a Novel, Hardware-Agnostic Pattern Recognition Algorithm}

\author{E Vourliotis$^{1a}$ and P Chang$^{2}$, P Elmer$^{3}$, Y Gu$^{1}$, J Guiang$^{1}$, V Krutelyov$^{1}$, B V Sathia Narayanan$^{1}$, G Niendorf$^{4}$, M Reid$^{4}$, M Silva$^{2}$, A Rios Tascon$^{3}$, M Tadel$^{1}$, P Wittich$^{4}$, A Yagil$^{1}$\\
on behalf of the CMS Collaboration}

\affil{$^1$University of California San Diego, CA, US}
\affil{$^2$University of Florida, FL, US}
\affil{$^3$Princeton University, NJ, US}
\affil{$^4$Cornell University, NY, US}

\email{$^a$emmanouil.vourliotis@cern.ch}

\begin{abstract}
Charged particle reconstruction is one the most computationally heavy components of the full event reconstruction of Large Hadron Collider (LHC) experiments.
Looking to the future, projections for the High Luminosity LHC (HL-LHC) indicate a superlinear growth for required computing resources for single-threaded CPU algorithms that surpass the computing resources that are expected to be available.
The combination of these facts creates the need for efficient and computationally performant pattern recognition algorithms that will be able to run in parallel and possibly on other hardware, such as GPUs, given that these become more and more available in LHC experiments and high-performance computing centres.
Line Segment Tracking (LST) is a novel such algorithm which has been developed to be fully parallelizable and hardware agnostic.
The latter is achieved through the usage of the Alpaka library.
The LST algorithm has been tested with the CMS central software as an external package and has been used in the context of the CMS HL-LHC High Level Trigger (HLT).
When employing LST for pattern recognition in the HLT tracking, the physics and timing performances are shown to improve with respect to the ones utilizing the current pattern recognition algorithms.
The latest results on the usage of the LST algorithm within the CMS HL-LHC HLT are presented, along with prospects for further improvements of the algorithm and its CMS central software integration.
\end{abstract}

\section{Motivation and the Line Segment Tracking Algorithm}

The High Luminosity Large Hadron Collider (HL-LHC) is the planned upgrade of the Large Hadron Collider (LHC) of CERN, with the target to collect data of proton-proton collisions corresponding to an integrated luminosity of more than 3000\fbinv~\cite{HLLHC}.
This can only be achieved by considerably enhancing the instantaneous luminosity, which, in turn, implies a drastic increase in the number of simultaneous collisions (pileup, PU).
Because of this, the computational complexity of event reconstruction is projected to exceed the available computing resources, especially for the highly combinatorial task of trajectory pattern recognition of charged particles.
This leads to both an increased timing, jeopardizing the ability to reconstruct the data at the desired rate, and an increased cost due to the higher demands for processing power.

\begin{figure}[!hbtp]
    \begin{center}
    \includegraphics[width=1.0\textwidth]{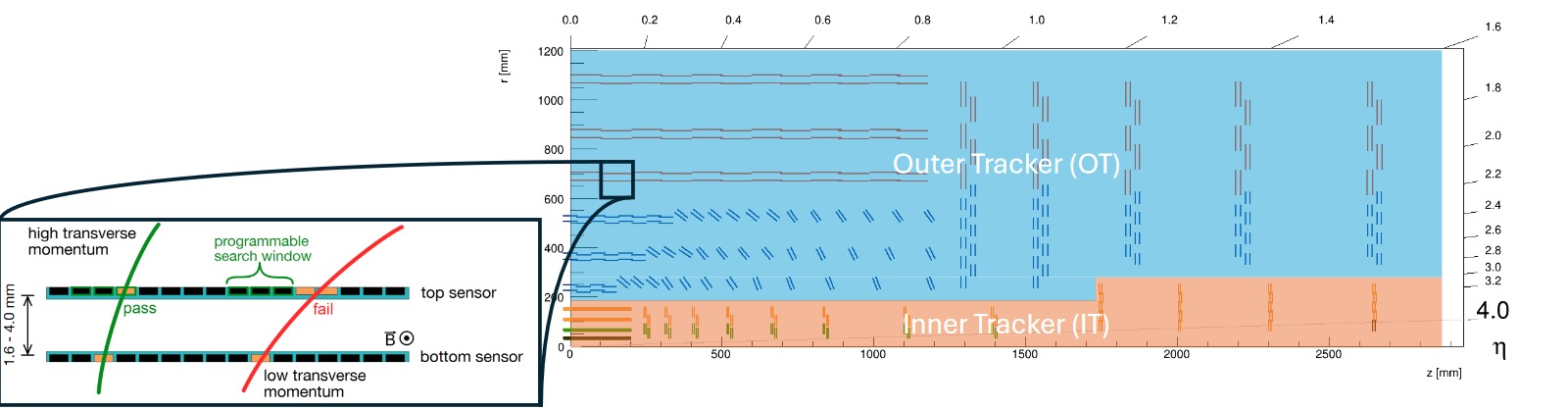}
    \end{center}
    \caption{A qualitative representation of the expected Phase-2 CMS tracker geometry~\cite{Phase2Tracker}.}
    \label{fig:phase2Tracker}
\end{figure}

To accommodate the HL-LHC conditions, the LHC experiments are planning a major upgrade of their software and hardware infrastructure (Phase 2).
The Line Segment Tracking (LST) algorithm aims at improving and parallelizing on Graphics Processing Units (GPUs) the charged hadron trajectory pattern recognition of the Phase 2 CMS experiment~\cite{CMSExperiment}.
It uses as input the hits of the CMS Phase 2 outer tracker (OT)~\cite{Phase2Tracker} and associates them to inner tracker (IT) tracks, ultimately producing a collection of OT+IT and OT-only track candidates.
The early stages of the algorithm rely to a significant degree on the characteristics of CMS Phase 2 OT, qualitatively shown in Fig.~\ref{fig:phase2Tracker}: one of the key aspects of its design is that each layer comprises of 2 closely-spaced silicon sensors.
In this way, two hits are recorded on each layer and are linked by the LST algorithm in a single pair of hits, called Mini Doublet (MD).
The MDs are useful in reducing the combinatorics for the trajectory patterns and have the advantage of being locally reconstructed, which is utilized by the LST algorithm to parallelize their creation.
Another handle for the reduction of the combinatorics is the definition of a lower \pT threshold for track reconstruction by tuning the search window for hit pairs.
The current lower \pT threshold for LST is set at 0.8\GeV.

MDs serve as the elementary building blocks for the LST algorithm to create tracks.
Based on precomputed connection maps for modules in the IT and OT, fulfilling geometric criteria that physical charged particle patterns obey, two MDs are linked to create a Line Segment (LS).
The selections used to create LST objects are described in Ref.~\cite{LSTLogic}, and incorporate also machine learning methods, as detailed in Ref.~\cite{LSTML}.
Two LSs with a common MD are subsequently linked to form a T3, and two T3s with a common MD are linked to form a T5.
T3s and T5s are combined with IT tracks (pLSs) to create pT3s and pT5s.
Since the reconstruction of the above objects only requires local information, it can be massively parallelized, as all objects of the same kind can be created concurrently. 
Out of those objects, only a subset is propagated to downstream algorithms to be made into full tracks:

\begin{itemize}[topsep=0pt,itemsep=0pt,parsep=0pt]
    \item pT5s, providing the majority of the efficiency.
    \item pT3s, complementing the pT5 efficiency.
    \item T5s, enabling the reconstruction of displaced tracks.
    \item Unlinked pLSs, covering the track reconstruction at high $|\eta|$, where there is no OT for the LST algorithm to create other objects.
\end{itemize}

\begin{figure}[!hbtp]
    \begin{center}
    \raisebox{0pt}{\begin{subfigure}{0.18\linewidth}
        \centering
        \includegraphics[width=1.0\textwidth]{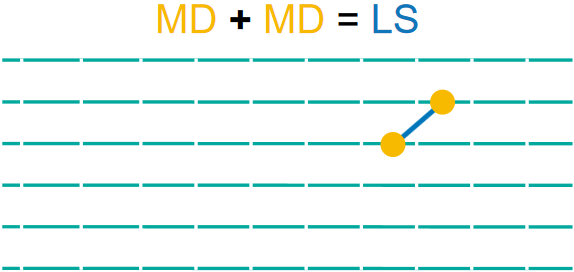}
        \captionsetup{skip=39pt}
        \caption{LS}
    \end{subfigure}}
    \raisebox{0pt}{\begin{subfigure}{0.18\textwidth}
        \includegraphics[width=1.0\textwidth]{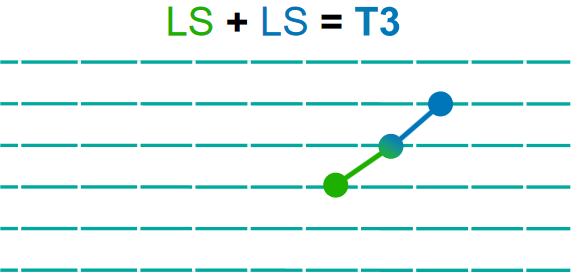}
        \captionsetup{skip=39pt}
        \caption{T3}
    \end{subfigure}}
    \raisebox{0pt}{\begin{subfigure}{0.18\textwidth}
        \includegraphics[width=1.0\textwidth]{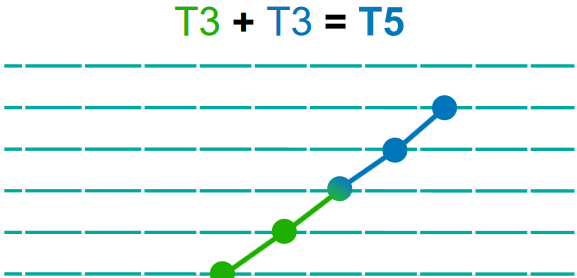}
        \captionsetup{skip=39pt}
        \caption{T5}
    \end{subfigure}}
    \begin{subfigure}{0.18\textwidth}
        \includegraphics[width=1.0\textwidth]{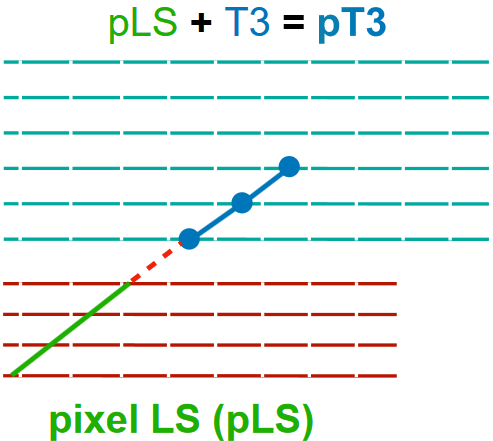}
        \caption{pT3}
    \end{subfigure}
    \begin{subfigure}{0.18\textwidth}
        \includegraphics[width=1.0\textwidth]{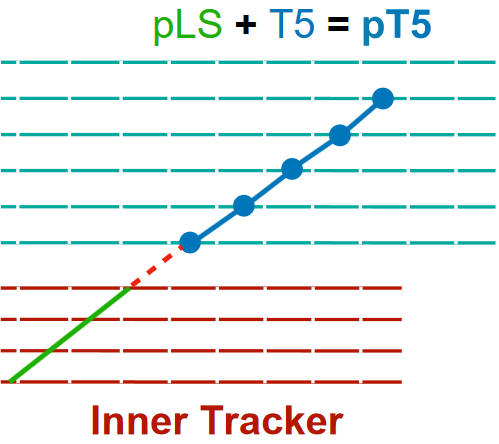}
        \caption{pT5}
    \end{subfigure}%
    \end{center}
    \caption{A qualitative representation of the different objects created by the LST algorithm~\cite{LSTHLT}.}
    \label{fig:LSTObjects}
\end{figure}

The objects created by the LST algorithm are summarized in Fig.~\ref{fig:LSTObjects}.
Previously, the LST algorithm performance had been demonstrated in the offline reconstruction setup in Ref.~\cite{LSTPerf}.
It is worth noting that the LST algorithm is written using the Alpaka abstraction framework~\cite{Alpaka1,Alpaka2,Alpaka3}, therefore it seamlessly runs on multiple hardware devices.

\section{Tracking in the Phase 2 CMS High Level Trigger}

The High Level Trigger (HLT) is one of the two tiers of the system that collect the events of interest for CMS.
It consists of a farm of processors running a version of the full event reconstruction software optimized for fast processing, and reduces the event rate before data storage.
As it is responsible for the acquisition of the data as they are produced by the (HL-)LHC, timing plays an important role for the algorithms it runs.
Apart from that, it needs to be general enough to cover for the majority of the potential physics signals.

The pattern recognition of tracks is of major importance for the HLT, as it is the basis of the reconstruction of most of the physics particles, while it needs to be run at a disproportionately short time for the combinatorial complexity of the task, especially at the harsh PU of HL-LHC.
The CMS Phase 2 HLT uses the Combinatorial Kalman Filter (CKF) to reconstruct tracks with $\pT > 0.9 \GeV$~\cite{HLTTDR}.
The track reconstruction is performed in two stages (iterations) based on different sets of initial track estimations (track seeding): the initial step produces tracks from pixel seeds with at least 4 hits (quads) created by the Patatrack algorithm~\cite{Patatrack1,Patatrack2}, while the highPtTriplet step produces tracks from pixel seeds with 3 hits (triplets), created by the legacy pixel seeding algorithm~\cite{TRK}.
It is worth noting that the CKF algorithm used for the trajectory pattern recognition (track building) in this configuration is inherently sequential, and is implemented on Central Processing Unit (CPU).
Once the built tracks, i.e. the set of hits originating from the same track, have been identified, these undergo a fitting procedure to extract the final track parameters, and are selected with requirements based on those parameters (tracking ID).
The ``highPurity" selection is applied, which provides a good balance between high efficiency and low fake+duplicate rate for prompt tracks~\cite{TRK}.
This baseline configuration will be mentioned below as ``Base CKF".

The LST algorithm can be an ideal candidate to run at the CMS Phase 2 HLT.
LST allows for the parallel processing of track reconstruction on GPUs, hence keeping the timing under control, while extending the physics acceptance of the HLT to displaced tracks.
In the following, a few potential configurations for integrating the LST algorithm in the CMS Phase 2 HLT are documented.
The LST algorithm is utilized as a replacement for track building for the initial step, using pixel seeds with at least 3 hits as pLSs.
Since the highPurity tracking ID has been optimized for prompt tracks, it is not applied to the LST objects targeting displaced tracks (T5s).
This leads to a high efficiency for displaced tracks, without any significant increase in the fake and duplicate rate.
Finally, as mentioned above, LST cannot build tracks for $|\eta| \gtrsim 2.5$, as that region is outside of the OT acceptance (Fig.~\ref{fig:phase2Tracker}).
As a result, the CKF algorithm still needs to run in the highPtTriplet step to recover efficiency in those high $|\eta|$ regions.
Different configurations are being used for the seeding of this ``recovery iteration": in the ``LST with CKF on Legacy Triplets" configuration, legacy triplets are used, in the ``LST with CKF on LST Quads" configuration only the quad LST pLSs are used, while in the ``LST with CKF on LST Quads+Triplets" configuration both the quad and triplet LST pLSs are used.
The last two configuration imply that the LST algorithm can be used not only as a track building but also as a track seeding algorithm.
All of the configurations described above are summarized in a more condensed format in Table~\ref{tab:configs}.

\begin{table}[htbp]
    \centering
    \caption{Summary of the HLT tracking sequence setup for each configuration described in this note~\cite{LSTHLT}.}
    \scalebox{0.8}{
    \begin{tabular}{ll|cccc}
        \multirow{2}{*}{Iteration} & \multirow{2}{*}{Procedure} & \multirow{2}{*}{Base CKF} & LST with CKF on & LST with CKF on & LST with CKF on  \\
         & & & Legacy Triplets & LST Quads & LST Quads+Triplets \\[3pt]
        \hline
        \multirow{2}{*}{} & \multirow{2}{*}{Seeding} & \multirow{2}{*}{Patatrack quads} & Patatrack quads + & Patatrack quads + & Patatrack quads + \\
         & & & Legacy Triplets & Legacy Triplets & Legacy Triplets \\[3pt]
        \multirow{2}{*}{Initial} & \multirow{2}{*}{Building} & \multirow{2}{*}{CKF} & \multirow{2}{*}{LST} & \multirow{2}{*}{LST} & \multirow{2}{*}{LST} \\
         & & & & & \\[3pt]
        \multirow{2}{*}{} & \multirow{2}{*}{Tracking ID} & \multirow{2}{*}{highPurity} & highPurity (pT3, pT5, pLS) & highPurity (pT3, pT5) & highPurity (pT3, pT5) \\
         & & & None (T5) & None (T5) & None (T5) \\[3pt]
        \hline
        \multirow{2}{*}{} & \multirow{2}{*}{Seeding} & \multirow{2}{*}{Legacy triplets} & \multirow{2}{*}{Legacy triplets} & \multirow{2}{*}{LST pLS quads} & LST pLS \\
         & & & & & quads+triplets \\[3pt]
        HighPt & \multirow{2}{*}{Building} & \multirow{2}{*}{CKF} & \multirow{2}{*}{CKF} & \multirow{2}{*}{CKF} & \multirow{2}{*}{CKF} \\
        Triplet & & & & & \\[3pt]
        \multirow{2}{*}{} & \multirow{2}{*}{Tracking ID} & \multirow{2}{*}{highPurity} & \multirow{2}{*}{highPurity} & \multirow{2}{*}{highPurity} & \multirow{2}{*}{highPurity} \\
         & & & & & \\[3pt]
    \end{tabular}
    }
    \label{tab:configs}
\end{table}

\section{Physics Performance and Throughput}

This section is dedicated to the measurement of the physics and computational performance of the CMS Phase 2 HLT configurations using LST for track reconstruction.
Three metrics are used for the physics performance:

\begin{itemize}[topsep=0pt,itemsep=0pt,parsep=0pt]
    \item Efficiency: The fraction of the matched simulated tracks from the hard-scattering vertex.
    \item Fake rate: The fraction of reconstructed tracks not matched to any simulated track.
    \item Duplicate rate: The fraction of reconstructed tracks matched to any simulated track that is matched to multiple reconstructed tracks.
\end{itemize}

For the measurement of the efficiency, the simulated tracks from the hard-scattering vertex are matched to the reconstructed tracks.
A given simulated track is considered a match to a reconstructed one if more than 75\% of the hits of the reconstructed track originate from the simulated track.
For the measurement of the fake and the duplicate rate, all simulated tracks are used for the matching.
In the following, any selections applied to the simulated or reconstructed tracks (depending on the metric, as described above) are shown on the plots.
The radial distance and the z position of the production vertex of the tracks are denoted as $\text{r}_\text{vertex}$ and $\text{z}_\text{vertex}$ respectively.
A simulated $t\bar{t}$ sample produced with 200 PU for the upgraded Phase 2 detector geometry is used for the measurements below.

\begin{figure}[!hbtp]
    \begin{center}
    \includegraphics[width=0.41\textwidth]{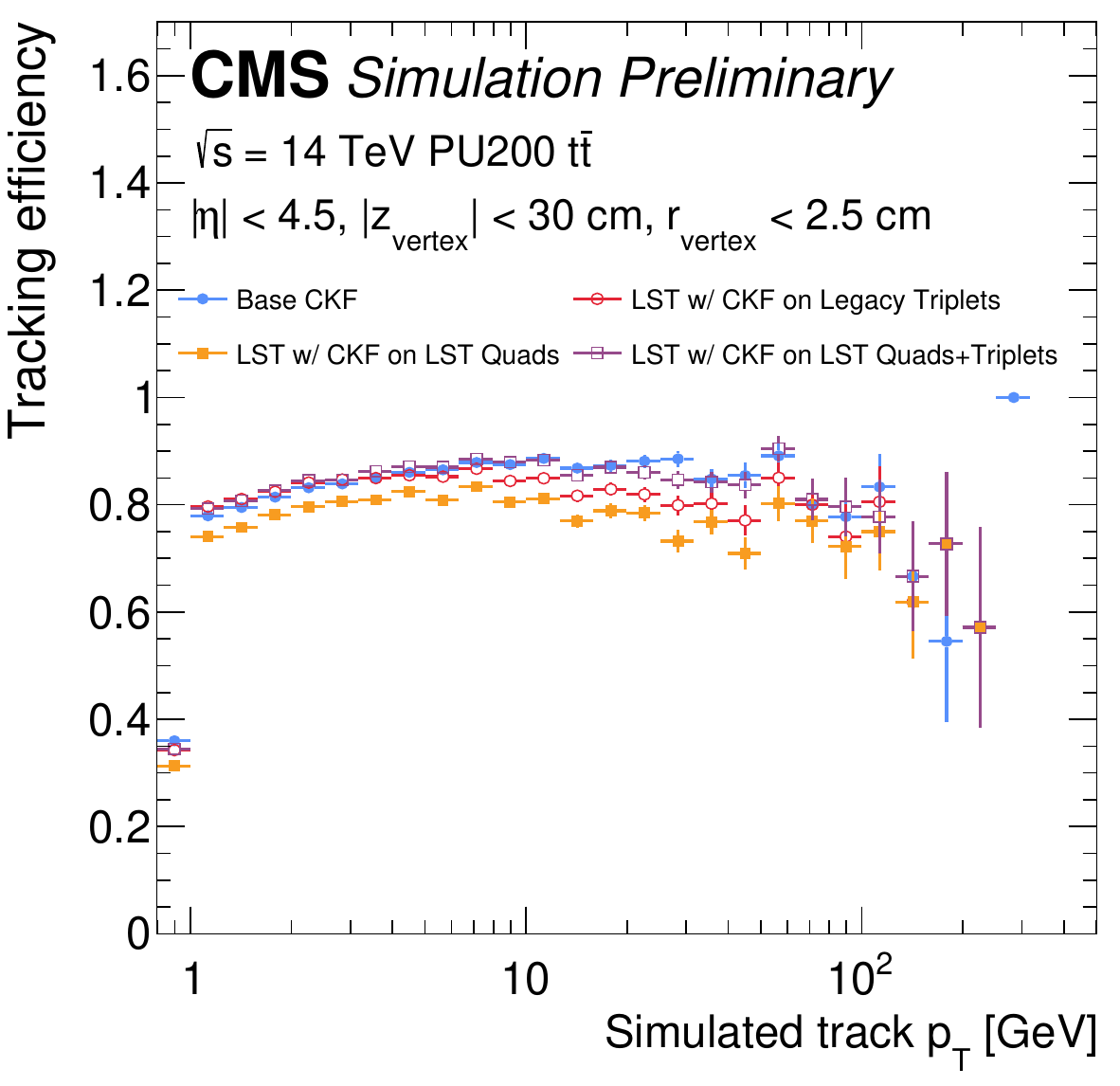}
    \includegraphics[width=0.41\textwidth]{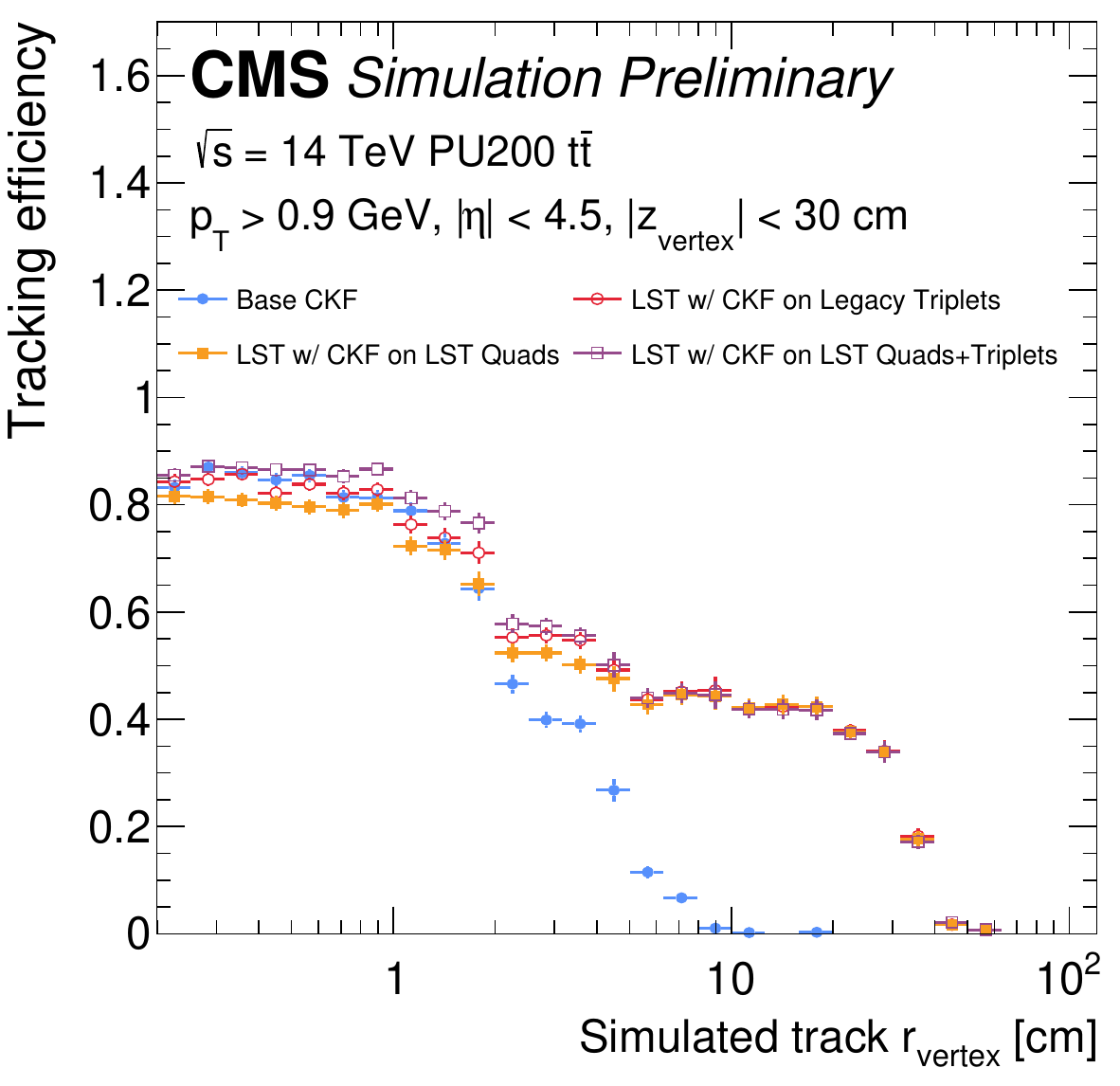}
    \end{center}
    \caption{The tracking efficiency is shown for Base CKF (blue), LST with CKF on Legacy Triplet (red), LST with CKF on LST Quads (orange) and LST with CKF on LST Quads+Triplets (purple) as a function of the simulated track \pT (left) and $\text{r}_\text{vertex}$ (right)~\cite{LSTHLT}.}
    \label{fig:eff_pt_dxy}
\end{figure}

Figure~\ref{fig:eff_pt_dxy} shows the efficiency of the different configurations tested as a function of the simulated track \pT (left) and $\text{r}_\text{vertex}$ (right).
It is obvious that the efficiency is lower when using only quads in the recovery iteration, highlighting the importance of triplets for track seeding in the current setup.
When triplets are used, the LST configurations reach an efficiency that is comparable to the one by Base CKF, or even higher for $\pT \lesssim 5 \GeV$.
The efficiency as a function of  $\text{r}_\text{vertex}$ demonstrates the fact that any configuration using LST for track building allows for acceptance of displaced tracks ($\text{r}_\text{vertex} \gtrsim 5~\text{cm}$).
This constitutes a completely new feature for the CMS HLT.
Notably, the efficiency drops at the radial distances roughly corresponding to tracker layers, with an endpoint at $\sim\!35~\text{cm}$, where less than 4 OT layers are available, so no T5s can be built.

\begin{figure}[!hbtp]
    \begin{center}
    \includegraphics[width=0.41\textwidth]{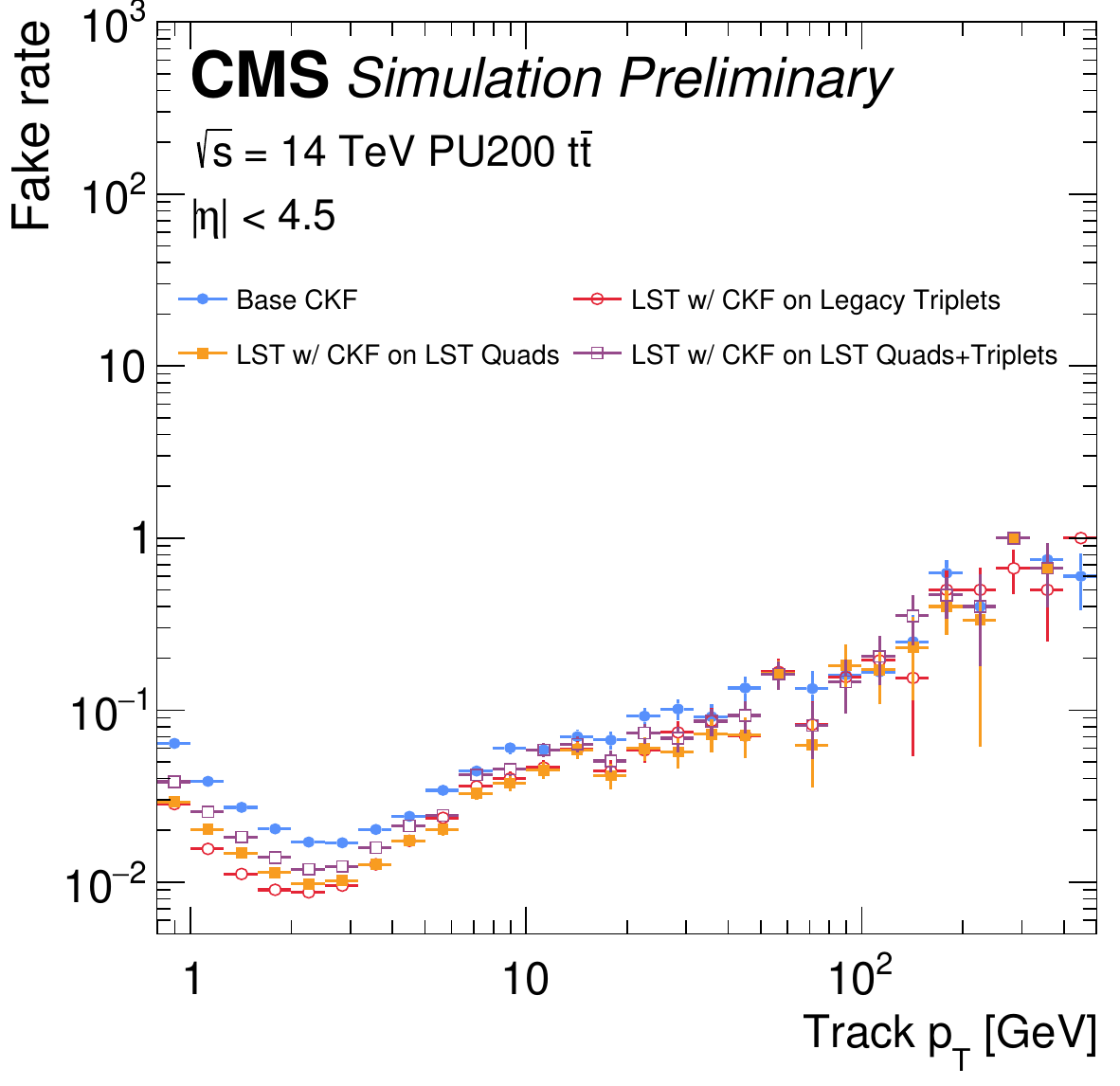}
    \includegraphics[width=0.41\textwidth]{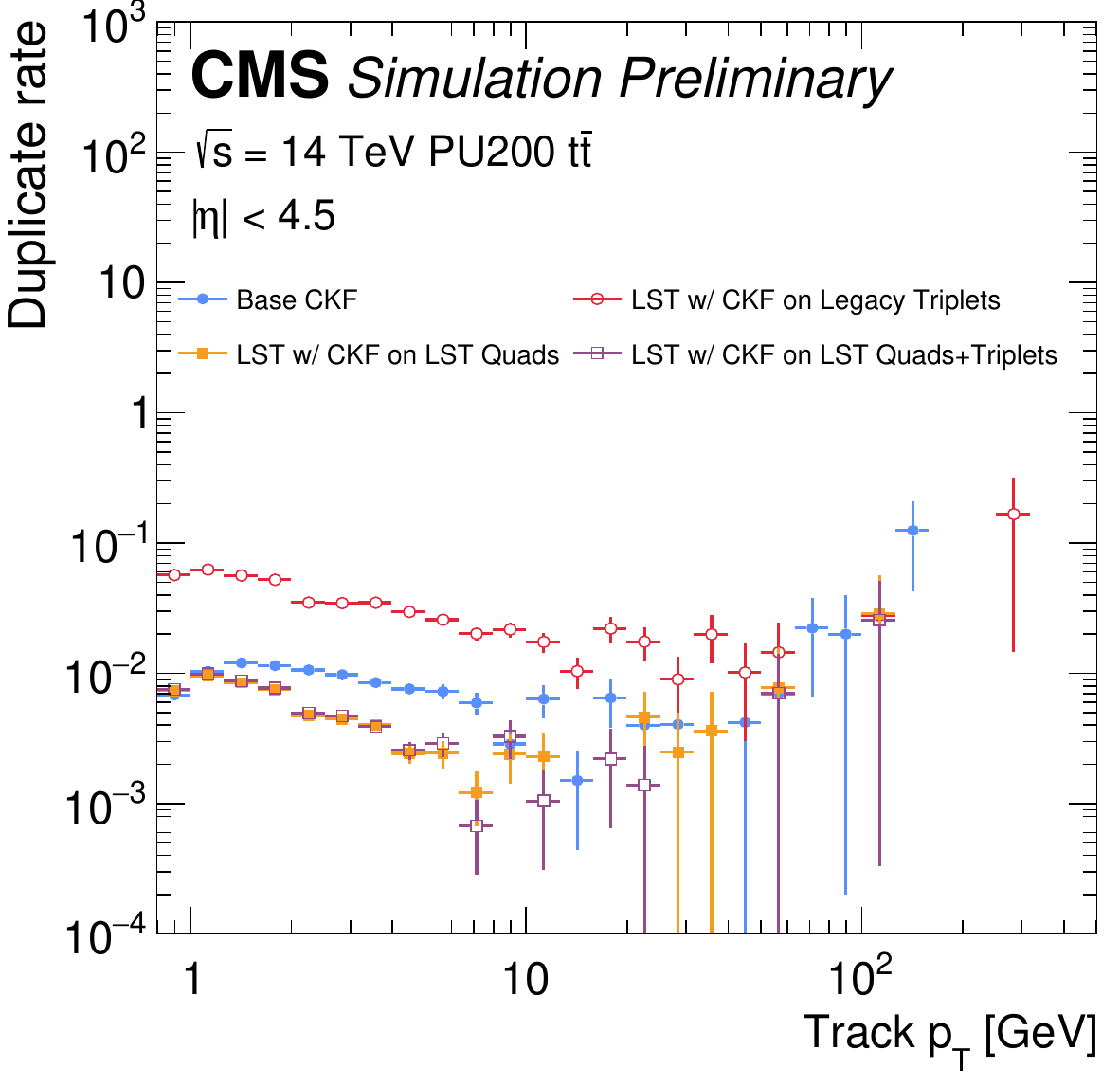}
    \end{center}
    \caption{The tracking fake rate (left) and duplicate rate (right) are shown for Base CKF (blue), LST with CKF on Legacy Triplet (red), LST with CKF on LST Quads (orange) and LST with CKF on LST Quads+Triplets (purple) as a function of the reconstructed track \pT~\cite{LSTHLT}.}
    \label{fig:fr_dr_pt}
\end{figure}

The fake rate is lower for any configuration using LST for track building.
Importantly, the right plot of Fig.~\ref{fig:fr_dr_pt} shows that most of the fake rate reduction comes for tracks with $\pT < 10 \GeV$.
Given that the bulk of tracks have low \pT, this implies a significant reduction of computational resources downstream, as less tracks need to be processed.
The left plot of Fig.~\ref{fig:fr_dr_pt} indicates a higher duplicate rate when the recovery CKF iterations runs on legacy triplets, as both legacy triplets and LST pLSs have the potential to reconstruct the same track.
When only the LST pLSs are used for track seeding, the overall duplicate rate is lower throughout the whole \pT range.

Based on the computational performance of the current offline tracking reconstruction, displaced tracking takes as much time as the prompt track reconstruction.
As displaced tracking is completely missing from the current CMS Phase 2 HLT configuration, its addition would imply a 50\% reduction of the throughput.
Table~\ref{tab:throughput} shows the throughput of the tracking sequence for CMS Phase 2 HLT configurations using LST, normalized to that of the Base CKF configuration.
The measurements were performed with 2 threads (for CPU), pinned to 2 specific CPU cores, and 2 streams (for GPU)  with local access to the input.
An AMD EPYC ``Milan" 7763 CPU and an NVIDIA ``Ampere" A30 PCIe GPU were used.
The results imply that a throughput reduction of at most 30\% is expected when running LST, hence including displaced tracking and improving various performance metrics, as outlined above.
The throughput reduction comes only for the LST configuration that run single-threaded on CPU.
When the LST configurations are executed on GPU, the throughput is comparable with the Base CKF one or even increases.
It was observed that most of the slowdown for the LST configurations is actually coming from the recovery iteration, which is run with CKF.

\begin{table}[!hbtp]
    \centering
    \caption{Throughput of the HLT tracking sequence for different configurations, normalized to that of the Base CKF one~\cite{LSTHLT}.}
    \begin{tabular}{l|ccc}
        \multirow{2}{*}{} & LST with CKF on & LST with CKF on & LST with CKF on  \\
         & Legacy Triplets & LST Quads & LST Quads+Triplets \\[3pt]
        \hline
        LST on CPU & \multirow{2}{*}{$0.72 \pm 0.07$} & \multirow{2}{*}{$0.86 \pm 0.07$} & \multirow{2}{*}{$0.70 \pm 0.09$} \\
        Throughput / Base CKF & & & \\[3pt]
        LST on GPU & \multirow{2}{*}{$1.03 \pm 0.09$} & \multirow{2}{*}{$1.35 \pm 0.12$} & \multirow{2}{*}{$0.92 \pm 0.09$} \\
        Throughput / Base CKF & & & \\[3pt]
    \end{tabular}
    \label{tab:throughput}
\end{table}

\section{Summary and Outlook}

LST is a novel, hardware-agnostic pattern recognition algorithm, targeting application to the CMS Phase 2 tracking.
The algorithm can bring improvements both to the physics performance, as it extends the acceptance of the current tracking implementation to displaced tracks, and to the computational performance, as it efficiently runs on GPUs, potentially increasing the event reconstruction throughput.
As such, it is a suitable candidate for performing the charged particle trajectory pattern recognition at the CMS Phase 2 HLT, where both the physics acceptance and timing considerations are of utmost importance.
This work presented the first exploratory integration of the LST algorithm in the CMS Phase 2 HLT, demonstrating a lot of potential for the future.

On top of the improvements showcased above, more developments are planned both for the LST algorithm and for the CMS Phase 2 HLT configuration.
The former involve the creation of more objects, the integration of more machine learning methods and the optimization of its implementation on CPU for LST, while the latter revolve around the optimization of the usage of the Patatrack algorithm for track seeding and the usage of the mkFit algorithm~\cite{mkFitPaper,mkFitDP} for the track building of the recovery iteration.

\section*{Acknowledgements} \label{sec:Acknowledgements}

This work was supported by the U.S. National Science Foundation under Cooperative Agreements OAC-
1836650, PHY-2323298, and PHY-2121686 and grant PHY-2209443.

\bibliographystyle{iopart-num}
\bibliography{LST_ACAT2024Proceedings}

\end{document}